\documentclass[conference]{IEEEtran}
\IEEEoverridecommandlockouts

\usepackage{listings}
\usepackage{amsmath}
\usepackage{amssymb}
\usepackage{algorithm}
\usepackage{algorithmic}
\usepackage{amsthm}
\usepackage{xcolor}
\usepackage{tabularx}
\usepackage{colortbl}
\usepackage{caption}
\usepackage{multirow}
\usepackage{makecell}
\usepackage[most]{tcolorbox}

\usepackage[
colorlinks=true,
linkcolor=blue,
citecolor=blue,
urlcolor=blue
]{hyperref}

\lstdefinestyle{acmstyle}{
    basicstyle=\ttfamily\small,
    breaklines=true,
    frame=single,
    columns=fullflexible,
    keepspaces=true,
    showstringspaces=false,
    backgroundcolor=\color{gray!8},
    rulecolor=\color{gray!50},
    xleftmargin=2pt,
    xrightmargin=2pt,
    aboveskip=6pt,
    belowskip=6pt
}

\usepackage{booktabs}
\usepackage{siunitx}
\usepackage{tabularray}
\usepackage{cite}
\usepackage{graphicx}
\usepackage{textcomp}
\usepackage{xcolor}
\usepackage{multirow}

\def\BibTeX{{\rm B\kern-.05em{\sc i\kern-.025em b}\kern-.08em
    T\kern-.1667em\lower.7ex\hbox{E}\kern-.125emX}}

\begin{document}

\title{Conventional Commit Classification using Large Language Models and Prompt Engineering\\
}

\author{
\IEEEauthorblockN{
Sakib Al Hasan\textsuperscript{*}\quad
H. M. Sazzad Quadir\quad
Md. Nurul Ahad Tawhid
}
\IEEEauthorblockA{
\textit{Institute of Information Technology} \\
\textit{University of Dhaka} \\
Dhaka, Bangladesh \\
bsse1209@iit.du.ac.bd \\
msse2015@iit.du.ac.bd \\
tawhid@iit.du.ac.bd
}
\thanks{\textsuperscript{*}Corresponding author: bsse1209@iit.du.ac.bd}
}

\maketitle

\begin{abstract}
Conventional commits provide a structured format for writing commit messages, which improves readability, software maintenance, and enables automation tools such as changelog generators and semantic versioning systems. Existing approaches to conventional commit classification typically rely on ML/DL models trained on large labeled datasets. In this paper, we investigated a training-free alternative by leveraging large language models (LLMs) through prompt engineering. Rather than building a task-specific classifier, we evaluate three prompting strategies, such as zero-shot, few-shot, and chain-of-thought, across three open-source LLMs of varying scale: Mistral-7B-Instruct, LLaMA-3-8B, and DeepSeek-R1-32B. Classification is performed directly on code diffs extracted from a balanced dataset of 3,200 commits mined from the InfluxDB repository, without any model fine-tuning. Our results show that few-shot prompting consistently achieves the highest accuracy, while chain-of-thought prompting does not yield additional gains for this classification task. Among the evaluated models, DeepSeek-R1-32B achieves the strongest overall performance, suggesting that model scale plays a meaningful role in conventional commit classification. These findings provide practical guidance for researchers and practitioners seeking to automate commit classification without the overhead of curating and maintaining labeled training data.
\end{abstract}


\section{Introduction}
Version control systems play a vital role in modern software development processes by providing tools that allow developers to track changes to their codebase, collaborate with others, and maintain a rich history. In the current software development landscape, Git has established itself as the \textit{de facto} version control tool, with commit messages serving as the primary means of documenting codebase changes. Well-structured commit messages improve the overall quality of the codebase, foster collaboration among software developers, and enable the use of automation tools such as changelog generators and release automation tools \cite{tian2022what}.

To improve the quality of commit messages and standardize them, the Conventional Commits specification was developed \cite{conventional_commits}. This specification provides a structured format for writing commit messages that typically include a specific type (e.g., \texttt{feat}, \texttt{fix}, \texttt{docs}, \texttt{style}, \texttt{refactor}, \texttt{test}, \texttt{chore}) of change to the codebase, along with a brief description of that change. This format allows software developers to leverage automation tools that perform semantic versioning and improve the traceability of changes. In practice, software developers often ignore the conventional commits specification and write unstructured commit messages that may need to be classified or converted to conventional commits \cite{zeng2025first}.

This automatic classification of commit messages into these predefined categories has thus emerged as a prominent research area in software engineering. Typically, prior studies rely on applying traditional machine learning and deep learning techniques on commit message datasets \cite{sarwar2020multi}, as well as natural language processing techniques to automatically create commit messages from diffs \cite{jiang2017automatically}. Although these techniques are known to achieve good performance, they require considerable data preprocessing, feature engineering, and maintenance as data and project terminology change.

Recent developments in LLMs have shown a promising direction in tackling various natural language understanding tasks without requiring any task-specific model \cite{hou2023large}. LLMs are known to perform complex tasks, such as classification, reasoning, and analysis, through carefully designed prompts, a concept commonly known as prompt engineering \cite{brown2020language, wei2022chain}. It allows researchers to leverage the knowledge embedded in LLMs without requiring any machine learning pipeline.

In this work, we investigate the effectiveness of LLMs for conventional commit classification tasks through the lens of prompt design. Rather than building a traditional machine learning pipeline, we construct prompts to guide an LLM in classifying commit messages according to the Conventional Commits specification. Since LLM outputs are inherently conditioned on prompt formulation, we treat model performance as an indicator of prompt effectiveness. We empirically evaluate this approach on a dataset constructed from the open-source GitHub repository InfluxDB to assess the accuracy and viability of the designed prompts.

The major contributions of the current work can be listed as follows:
\begin{itemize}
    \item This paper proposes a new approach to conventional commit classification using LLMs and prompt engineering instead of traditional machine learning.
    \item This paper provides an empirical evaluation of the suggested approach with a broad set of data.
    \item This paper discusses the effectiveness of prompt-based classification and identifies the benefits and disadvantages of the approach within the framework of software engineering.
\end{itemize}

The rest of the paper is structured as follows:
Section \ref{bg} describes the background on the analysis and classification of commit messages, Section \ref{methodology} describes the suggested approach, Section \ref{result} describes the experiment and provides the results, Section \ref{discussion} describes the discussion, and Section \ref{conclusion} provides the concluding remarks and future directions.

\section{Background}
\label{bg}

This section provides the conceptual foundation for the present study. We begin with an overview of the Conventional Commits specification and its role in standardizing commit messages, followed by a review of existing approaches to automated commit message classification. We then discuss the emergence of large language models and prompt engineering as a training-free alternative for classification tasks. Finally, we survey the two studies most directly related to our work and identify the gaps that motivate the approach proposed in this paper.

\subsection{Conventional Commits}
Commit messages are an essential component of version control systems, providing critical context about the changes introduced to a codebase. However, when these messages are written in an inconsistent and unstructured manner, it becomes difficult to automate downstream tasks such as release management, changelog generation, and version tracking \cite{tian2022what}. To address this issue, the Conventional Commits specification was introduced as a lightweight convention for standardizing commit messages \cite{conventional_commits}.

The specification dictates that each commit must begin with a type indicating the nature of the change, optionally followed by a scope, and finally a short description. The general syntax is:
\begin{center}
    \texttt{<type>[optional scope]: <description>}
\end{center}

Common commit types include:
\begin{itemize}
    \item \texttt{feat}: Introduces a new feature.
    \item \texttt{fix}: Resolves a bug or issue.
    \item \texttt{docs}: Changes related strictly to documentation.
    \item \texttt{style}: Code style changes that do not affect functionality (e.g., formatting).
    \item \texttt{refactor}: Code changes that neither fix a bug nor add a feature.
    \item \texttt{test}: Additions or modifications to testing suites.
    \item \texttt{chore}: Routine maintenance tasks or build-related changes.
\end{itemize}

Adherence to this structural design allows developers to maintain a highly organized commit history. Additionally, tools, including changelog generators and semantic versioning tools, inherently rely on these data structures to determine the appropriate version increment of the software \cite{zeng2025first}. Despite the aforementioned advantages, adherence to the data structures is not consistent across all systems. Therefore, determining the appropriate type of commit for unstructured legacy systems often requires manual inspection and/or automated classification tools.

\subsection{Commit Message Classification}
Commit message classification is a research area within software engineering that deals with the classification of commit messages into predefined semantic groups based on the textual information they contain. Various research studies have been carried out to automate the commit message classification process by employing machine learning techniques\cite{jiang2017automatically, sarwar2020multi}. These techniques involve the training of machine learning models on a dataset with predefined classes to classify the commit messages into the predefined classes \cite{sarwar2020multi}. These machine learning pipelines involve a series of complex pre-processing steps followed by the training of the machine learning model on the dataset with the help of various machine learning algorithms like SVM, Random Forests, etc. \cite{sarwar2020multi}. Similar research areas like the generation of commit messages from diff also involve the training of complex neural machine translation models \cite{jiang2017automatically}.

Although the classification system employing machine learning algorithms performs with high accuracy, it also poses a series of logistical issues. Building an effective machine learning model for the commit message classification system requires a dataset with a large number of examples belonging to predefined classes. Moreover, the machine learning model also needs to be frequently updated by training the model on the newly acquired dataset or the updated classification taxonomy.

\subsection{Large Language Models and Prompt Engineering}
Recently, LLMs have shown unprecedented performance in natural language understanding and zero-shot reasoning tasks. LLMs are trained on a huge variety of text data and are capable of performing a variety of tasks such as summarization, question answering, and classification, among others, without requiring task-specific fine-tuning \cite{brown2020language}.

A highly effective method of adapting LLMs to a specific task is prompt engineering, where a prompt is used to guide the model in generating a response to a question. Rather than building a classifier from scratch, a prompt or instruction is used to guide the language model in classifying a given input text into a predefined category \cite{brown2020language, wei2022chain}.

This approach provides a number of advantages, including a significant reduction in the need to collect a huge set of labeled data and train a machine learning model on it. Prompt tuning or engineering also provides a high degree of flexibility in adapting a classification rule by simply adjusting the prompt. LLMs also have a high level of understanding of software engineering terminology, which could potentially improve their performance in comparison to a frequency-based approach \cite{hou2023large}.

While these models have increasingly been adopted in software engineering for tasks such as code generation, bug detection, and documentation generation, their specific application to commit message classification—particularly within the rigorous framework of the Conventional Commits specification—remains an emerging area that warrants comprehensive empirical investigation \cite{zeng2025first}.

\subsection{Related Work}
Commit classification has been studied extensively in the software engineering literature. Early work relied on keyword-based analysis of commit messages and hand-crafted features to distinguish between maintenance activity categories~\cite{swanson1976dimensions}. Hindle et al.~\cite{hindle2009automatic} applied machine learning to automatically classify commits into corrective, adaptive, and perfective categories, establishing a benchmark that subsequent studies built upon. Levin and Yehudai~\cite{levin2017boosting} extended this by combining commit message keywords with source code change metrics to improve classification accuracy. On the other hand, Sarwar et al.~\cite{sarwar2020multi} employed transfer learning on commit message text to perform multi-label classification across maintenance categories. Although these approaches achieve reasonable accuracy, they depend heavily on labeled training data and require retraining whenever the classification taxonomy or project context changes.

More recent work has shifted towards leveraging pre-trained language models to improve classification quality without extensive feature engineering. Ghadhab et al.~\cite{ghadhab2021augmenting} combined BERT with fine-grained source code change features extracted at the method and statement level, demonstrating that incorporating code-level signals alongside commit messages yields consistent and significant gains over text-only approaches. Similarly, Jiang et al.~\cite{jiang2017automatically} showed that neural models trained on code diffs can capture the semantic intent of a change, further reinforcing the value of diff-based input for commit understanding tasks.

Zeng et al.~\cite{zeng2025first} extended this direction to the Conventional Commits Specification by fine-tuning a CodeLlama-7B model on a manually annotated dataset of 2,000 commits mined from 116 open-source repositories, but their approach relies on supervised fine-tuning on domain-specific labeled data and evaluates LLMs under only a single fixed prompting configuration, leaving the influence of prompt design on CCS classification performance entirely unexplored. Sazid et al.~\cite{sazid2024commit} applied in-context learning with GPT-3 to commit classification in a training-free setting and evaluated zero-shot, one-shot, and few-shot configurations, but their study operates on commit message text alone without incorporating code diff information, evaluates only a single proprietary model without considering how behavior varies across model architectures and parameter scales, and targets the coarser three-category Swanson taxonomy rather than the Conventional Commits specification. The present work addresses these gaps by systematically evaluating multiple state-of-the-art prompting strategies across open-source LLMs of varying scale using code diffs as the primary input signal, without requiring any labeled training data.

\section{Methodology}
\label{methodology}

\begin{figure*}[t]
\centering
\includegraphics[width=0.7\textwidth]{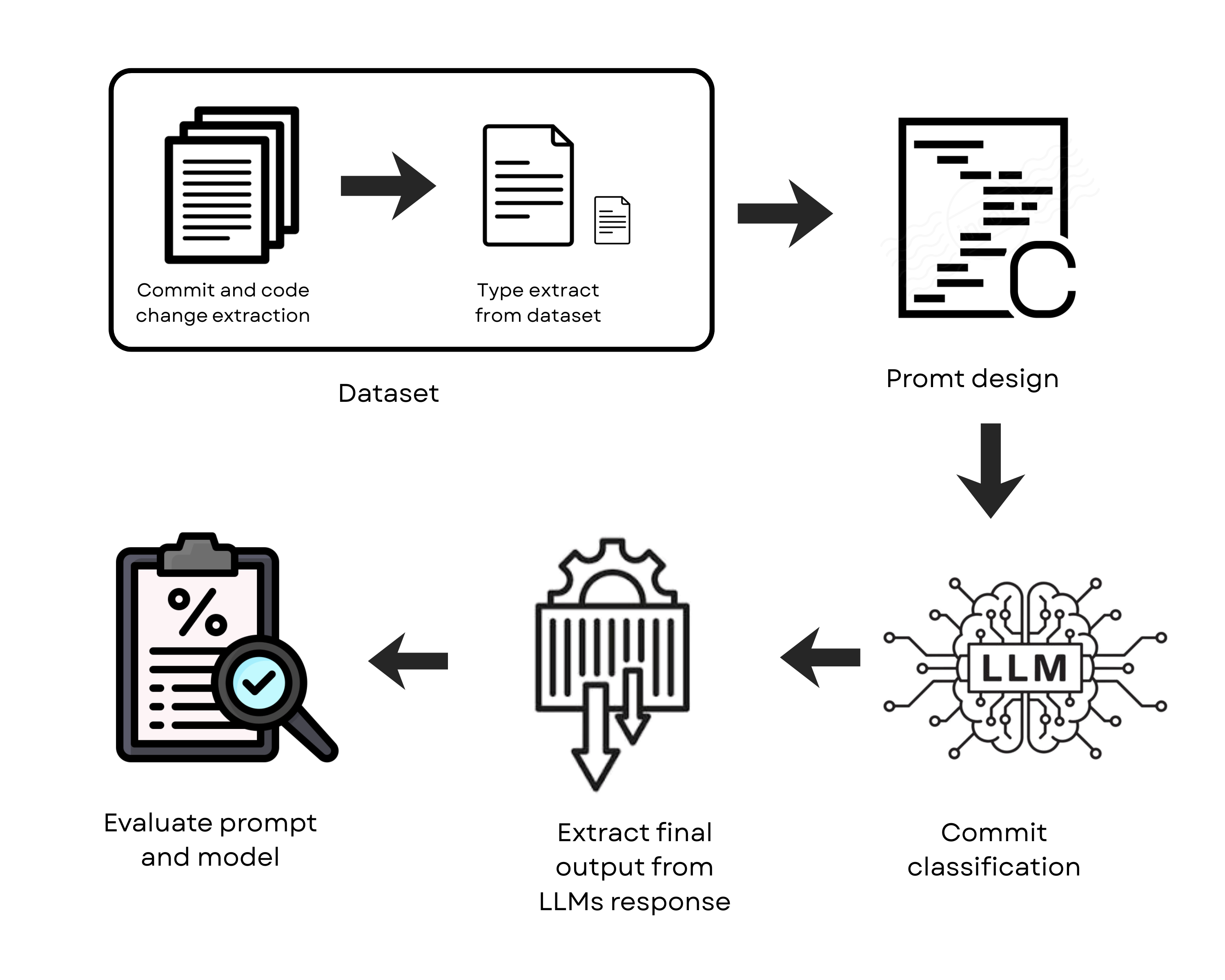}
\caption{Overview of the proposed methodology.}
\label{fig:methodology}
\end{figure*}

This section describes the experimental methodology used to evaluate commit classification using LLM. The methodology consists of collecting a dataset from a public GitHub repository describe in subsection A, extracting commits and code changes, and selecting 3200 commits and code changes to evaluate prompt and LLM performance. The overall goal is to check that an LLM can classify the conventional commit based on the code change.

Figure~\ref{fig:methodology} presents the overall workflow of the proposed framework. The process consists of five main stages: dataset preparation, prompt design, applying the prompt and codupdating the diff. code to the LLM, comparing the LLM's classification to the ground truth, and evaluating the model and the prompt. Below, we explain each of the stages separately with necessary details:

\subsection{Dataset collection}

\begin{figure}[h]
    \centering
    \includegraphics[width=0.9\linewidth]{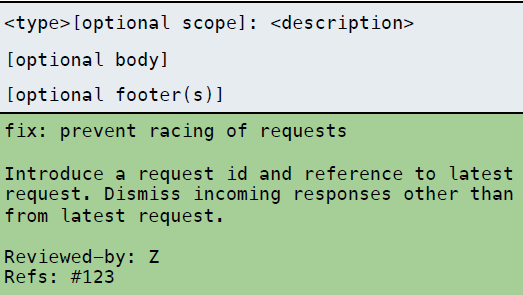}
    \caption{Conventional commit format and an example}
    \label{fig:commit}
\end{figure}

We construct our dataset from the open-source GitHub repository InfluxDB\footnote{https://github.com/influxdata/influxdb}, which is a widely used Time Series Database (TSDB) project. The repository is highly active and popular, with over 31,500 stars, 3,700 forks, and approximately 50,000 commits contributed by a large number of developers. It spans multiple programming languages (e.g., Go, Python, and shell scripts) and covers diverse domains such as database systems, distributed systems, and cloud infrastructure, making it representative of real-world software engineering practices.
From this repository, we extract commit messages and their corresponding code changes. The commit structure is illustrated in Figure~\ref{fig:commit}. Each commit consists of two mandatory components—the commit type (which serves as the classification label) and the commit message—as well as optional components such as the body and footer, which provide additional contextual information about the changes.
In total, we collect 3,200 samples which is approximately balanced dataset by ensuring that each class contains a nearly equal number of samples, with only minor variations. No significant portion of the data was discarded during this process; instead, we preserved diversity while maintaining approximate balance across classes.

\subsection{Prompt Design}
There are different types of prompts, but we use three here: zero-shot, few-shot, and chain-of-thought prompting \cite{brown2020language, wei2022chain}. This prompting is popular and gives better results in previous studies\cite{white2023prompt}. In zero-shot, we just give the class overview and instructions on what he needs to classify. In a few shots, we give examples of each type along with the previous zero-shot prompting instruction. In chain-of-thought (CoT) prompting \cite{wei2022chain}, the model is encouraged to generate intermediate reasoning steps before producing the final classification. In this context, `reasoning' refers to a brief, step-by-step explanation in which the model identifies relevant cues from the commit message and code diff (e.g., keywords and intent), interprets their meaning, and maps them to an appropriate commit type. After generating this reasoning, the model outputs the final classification in the required format. For example, the model is instructed to first generate a brief step-by-step reasoning explaining which commit classification is appropriate and why, and then output the final answer in the required format. After that reasoning write the final output. Zero-shot prompt, like unsupervised classification, is based solely on each type's description and classifies the commit. On the other hand, few-shot prompting is similar to a supervised classification setting, where the model is provided with labeled examples to guide its predictions. In our prompt design, we include representative examples from each commit class to help the model learn the mapping between commit messages, code changes, and their corresponding labels. These examples are selected from the dataset to cover diverse commit patterns and avoid bias toward specific types of changes. Rather than purely random selection, we ensure that the examples are clear, unambiguous, and representative of their respective classes.

\begin{figure}[h]
    \centering
    \includegraphics[width=0.9\linewidth]{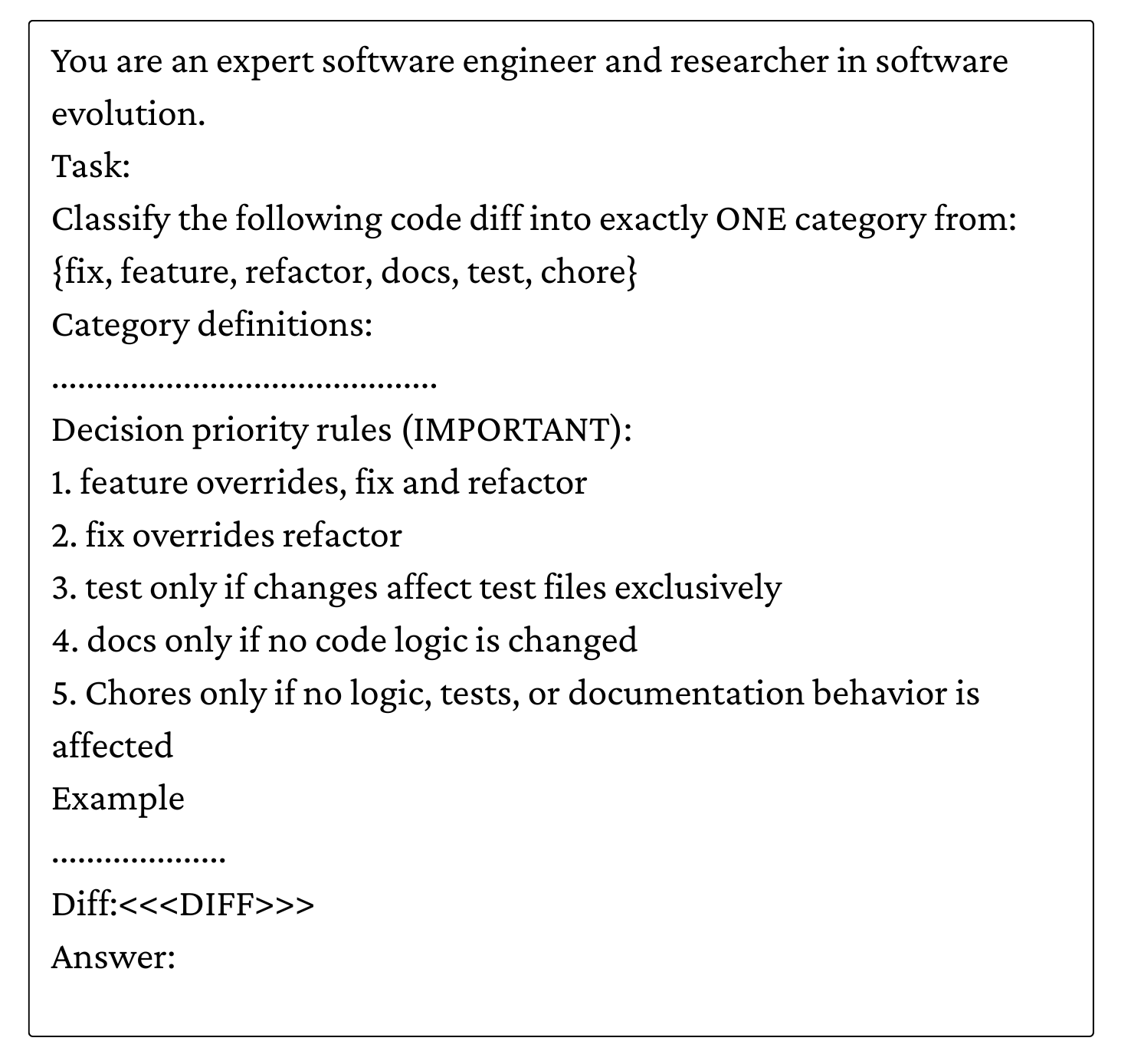}
    \caption{Prompt design example}
    \label{fig:promt}
\end{figure}

We ask the LLM to act as a software engineer and follow the provided instructions. The number and structure of the examples were determined empirically to balance prompt length and model performance. Additionally, we incorporate decision-priority rules, as shown in Figure \ref{fig:promt}, to guide the model when multiple classification cues are present. These rules were designed based on the Conventional Commits specification and common patterns observed in the dataset, such as prioritizing explicit commit type keywords (e.g., `fix', `feat') over inferred semantic meaning. Overall, the prompt design is guided by both the specification and observed commit behaviors in the dataset. 

\subsection{Commit Classification}
Once the prompts are designed, the code diff. and prompts are provided as input to the evaluated large language models. The models process the prompt and generate a predicted label corresponding to the commit type. This step represents the core classification process, in which the semantic understanding of the language models is applied to determine the intent of each code change. The model's classification output is recorded for further analysis.

After the classification process, the models' predicted commit labels are compared with the ground-truth labels in the dataset. This comparison allows us to determine whether each prediction is correct or incorrect. The results of this comparison are used to compute standard classification metrics such as accuracy, precision, recall, and F1-score. This step is essential for quantitatively evaluating how well the models perform in identifying the correct commit categories.

The overall process is described algorithmically in Algorithm \ref{alg:commit_classification}.

\begin{algorithm}[t]
\caption{LLM-Based Conventional Commit Classification}
\label{alg:commit_classification}
\begin{algorithmic}[1]

\REQUIRE Code change dataset $C$, set of models $L$, prompting strategies $S$
\ENSURE Evaluation metrics $Acc$, $Prec$, $Rec$, $F1$

\FOR{each model $l \in L$}
    \FOR{each prompting strategy $s \in S$}

        \STATE Initialize prediction set $P \leftarrow \emptyset$

        \FOR{each code change $c_i \in C$}

            \STATE Generate prompt according to strategy $s$
            
            \IF{$s =$ zero-shot}
                \STATE Construct instruction-based prompt
            \ELSIF{$s =$ few-shot}
                \STATE Include labeled code change examples
            \ELSIF{$s =$ chain-of-thought}
                \STATE Add step-by-step reasoning instruction
            \ENDIF

            \STATE Feed prompt to model $l$

            \STATE Obtain predicted label
            \STATE $\hat{y_i} \leftarrow \text{LLM}(prompt)$

            \STATE Store prediction
            \STATE $P \leftarrow P \cup \{\hat{y_i}\}$

        \ENDFOR

        \STATE Compute classification metrics

        \STATE Store accuracy, precision, recall and F1 score for model $l$ and strategy $s$

    \ENDFOR
\ENDFOR

\RETURN Evaluation metrics for all $(l,s)$ combinations

\end{algorithmic}
\end{algorithm}

\subsection{Evaluated Models}

To classify code change types, we evaluate three large-scale open-source large language models. They are Mistral-7B-Instruct \cite{jiang2023mistral}, LLaMA-3-8B \cite{ai2024llama}, and DeepSeek-R1-32B \cite{deepseek2025deepseek}. These models are selected because they represent different model architectures and parameter scales, allowing us to investigate how model size and design influence commit classification performance.

Mistral-7B-Instruct is a 7-billion-parameter instruction-tuned model designed for effective inference and robust performance on challenges involving general language comprehension. The LLaMA-3-8B model has a parameter size of 8 billion, and it was developed by Meta and optimized for conversational and general-purpose NLP tasks. The DeepSeek-R1-32B model has 32 billion parameters, which is a significantly larger model that enhances reasoning capabilities. By comparing these three models with varying parameter sizes, we are able to analyze how model capacity affects the ability to capture semantic patterns in change code and commit messages. We use both small- and large-parameter-size models to determine which perform better. In Table \ref{tab:evaluated_models}, we represent the model name and its parameter size.

\begin{table}[h]
\centering
\caption{Evaluated LLMs and parameter sizes}
\begin{tabular}{lc}
\toprule
Model & Parameters \\
\midrule
Mistral-7B-Instruct & 7B \\
LLaMA-3-8B & 8B \\
DeepSeek-R1-32B & 32B \\
\bottomrule
\end{tabular}
\label{tab:evaluated_models}
\end{table}

\subsection{Evaluation Metrics}

To measure the effectiveness of LLM models and prompts, we use standard classification metrics commonly used in machine learning and software engineering research. These metrics measure how accurately the models classify commit messages into their correct categories and provide insights into both prediction correctness and error distribution.

Precision measures the proportion of correctly predicted positive instances among all predicted positive instances. Recall evaluates the proportion of correctly identified positive instances among all actual positive instances in the dataset. The F1-score is the harmonic mean of precision and recall and provides a balanced measure of model performance when there is an imbalance between false positives and false negatives. In addition to these metrics, overall accuracy is reported as the proportion of correctly classified commits across the entire dataset.

The metrics that we used in this study are calculated as follows:

\begin{equation}
\text{Accuracy} = \frac{TP + TN}{TP + TN + FP + FN}
\end{equation}

\begin{equation}
\text{Precision} = \frac{TP}{TP + FP}
\end{equation}

\begin{equation}
\text{Recall} = \frac{TP}{TP + FN}
\end{equation}

\begin{equation}
F_1 = 2 \times \frac{\text{Precision} \times \text{Recall}}{\text{Precision} + \text{Recall}}
\end{equation}

where $TP$ denotes true positives, $FP$ denotes false positives, and $FN$ denotes false negatives.

\section{Result}
\label{result}

Table~\ref{tab:performance_results} presents the classification performance of the evaluated models under different prompting strategies. The results are reported using accuracy, precision, recall, and F1-score.

\begin{table*}[t]
\centering
\caption{Classification performance across models and prompting strategies}
\label{tab:performance_results}
\small
\setlength{\tabcolsep}{5pt}
\begin{tabular}{llcccc}
\toprule
\textbf{Model} & \textbf{Prompting Strategy} & \textbf{Accuracy} & \textbf{Precision} & \textbf{Recall} & \textbf{F1-score} \\
\midrule
\multirow{3}{*}{Mistral-7B-Instruct}
& Zero-Shot        & 0.3356 & 0.5044 & 0.3356 & 0.2824 \\
& Few-Shot         & \textbf{0.6154} & 0.3823 & \textbf{0.6154} & 0.4706 \\
& Chain-of-Thought & 0.5319 & 0.3032 & 0.5319 & 0.3842 \\
\midrule
\multirow{3}{*}{LLaMA-3-8B}
& Zero-Shot        & 0.3918 & 0.4495 & 0.3918 & 0.3897 \\
& Few-Shot         & 0.4012 & 0.4511 & 0.4012 & 0.3862 \\
& Chain-of-Thought & 0.3229 & 0.3777 & 0.3229 & 0.3206 \\
\midrule
\multirow{3}{*}{DeepSeek-R1-32B}
& Zero-Shot        & 0.5652 & 0.5703 & 0.5652 & 0.5260 \\
& Few-Shot         & 0.5749 & 0.5933 & 0.5749 & \textbf{0.5466} \\
& Chain-of-Thought & 0.5497 & \textbf{0.5966} & 0.5497 & 0.5402 \\
\bottomrule
\end{tabular}
\end{table*}

Across all models, the few-shot prompting strategy achieves the highest accuracy for Mistral-7B-Instruct (0.6154) and DeepSeek-R1-32B (0.5749), while LLaMA-3-8B achieves its highest accuracy with few-shot prompting (0.4012). In contrast, the lowest accuracy values are observed for zero-shot prompting in Mistral-7B-Instruct (0.3356) and for chain-of-thought prompting in LLaMA-3-8B (0.3229).

Table~\ref{tab:prompt_performance} summarizes the average performance of each prompting strategy aggregated across all models.

\begin{table}[t]
\centering
\caption{Average performance comparison of prompting strategies across three models}
\label{tab:prompt_performance}
\begin{tabular}{lccc}
\hline
\textbf{Metric} & \textbf{Zero-shot} & \textbf{Few-shot} & \textbf{Chain-of-Thought} \\
\hline
Accuracy  & 0.431 & 0.531 & 0.468 \\
Precision & 0.508 & 0.476 & 0.426 \\
Recall    & 0.431 & 0.531 & 0.468 \\
F1-score  & 0.399 & 0.468 & 0.415 \\
\hline
\end{tabular}
\end{table}

The aggregated results show that few-shot prompting achieves the highest average accuracy (0.531) and F1-score (0.468), while zero-shot prompting achieves the highest precision (0.508). Chain-of-thought prompting produces intermediate performance across all metrics.

Table~\ref{tab:model_performance} reports the average performance of each model across all prompting strategies.

\begin{table}[t]
\centering
\caption{Average performance of LLM models across different prompting strategies for commit classification}
\label{tab:model_performance}
\begin{tabular}{lccc}
\hline
\textbf{Metric} & \textbf{Mistral-7B-Instruct} & \textbf{LLaMA-3-8B} & \textbf{DeepSeek-R1-32B} \\
\hline
Accuracy  & 0.494 & 0.372 & 0.563 \\
Precision & 0.397 & 0.426 & 0.587 \\
Recall    & 0.494 & 0.372 & 0.563 \\
F1-score  & 0.379 & 0.366 & 0.538 \\
\hline
\end{tabular}
\end{table}

Among the evaluated models, DeepSeek-R1-32B achieves the highest average performance across all metrics, while LLaMA-3-8B records the lowest performance. Mistral-7B-Instruct shows intermediate performance across all metrics.

\section{Discussion}
\label{discussion}

A key finding of this study is that prompt design has a significant impact on classification performance, often exceeding that of the model itself. In particular, few-shot prompting consistently outperforms both zero-shot and chain-of-thought strategies. This suggests that LLMs benefit strongly from exposure to representative examples, which help them align input patterns with expected outputs.

From a practical perspective, this indicates that effective prompt design for classification tasks should prioritize the inclusion of clear, diverse, and representative examples. These examples act as implicit training signals that guide the model's behavior without requiring fine-tuning. In contrast, zero-shot prompting lacks this contextual grounding, leading to lower overall performance despite its simplicity.

Interestingly, chain-of-thought prompting does not improve results in this setting. This suggests that classification tasks, particularly those involving structured labels such as commit types, rely more on pattern recognition than on multi-step reasoning. As a result, adding explicit reasoning steps may introduce unnecessary complexity rather than improving accuracy.

\subsection*{Key Observations}

Based on the experimental results, the following key observations can be made:

\begin{itemize}
\item Few-shot prompting results in the most reliable improvement for commit classification tasks.
\item Chain-of-thought prompting does not necessarily improve performance for simple classification tasks.
\item Larger language models often achieve better performance due to their superior ability to comprehend context.
\item Prompting plays a crucial role in adapting general-purpose LLMs for software engineering tasks.
\end{itemize}

These observations provide useful guidelines for researchers who wish to leverage the power of large language models for software repository mining tasks.

\subsection*{Limitation}
However, the study is limited in some aspects, including the scope of the dataset used in the study, the prompts used in the study, and the models used in the study. In future research, the study can be extended in different ways. First, using more datasets from different software repositories can help us understand the generalization of the study's results. Second, applying different automated prompt optimization techniques can improve the performance of conventional commit message classification with large language models. Third, using a combination of large language models and other techniques like machine learning can help us propose a new approach that balances efficiency and effectiveness in conventional commit message classification.

While prior work has explored traditional machine learning and deep learning approaches for commit classification, this study does not aim to directly compare against or outperform these methods. Instead, our focus is on analyzing how different prompt designs influence LLM-based classification performance. By evaluating multiple prompting strategies, we aim to provide practical insights and guidelines on how prompt construction affects classification behavior and can be leveraged to improve results. As such, the performance comparisons in this study should be interpreted in a relative sense (i.e., across prompting strategies and models) rather than as evidence of superiority over existing approaches. Claims regarding improved performance of larger models refer only to comparisons within the evaluated LLMs and not to external baselines. A direct comparison with traditional ML/DL approaches is outside the scope of this work and is left for future research.

\section{Conclusion}
\label{conclusion}

In this study, we describe an empirical investigation into the effectiveness of large language models in conventional commit message classification using different prompting strategies. We investigate the effect of different prompting strategies on the performance of conventional commit message classification using three open-source language models: Mistral-7B-Instruct, LLaMA-3-8B, and DeepSeek-R1-32B. Our results demonstrate that different prompting strategies significantly affect the performance of conventional commit message classification using large language models. Our results also demonstrate that few-shot prompting is more promising compared to zero-shot prompting and chain-of-thought prompting. In addition, we found that larger models perform better at conventional commit message classification than smaller ones.

Finally, future research can be directed towards exploring the applications of conventional commit message classification in software engineering tasks like automated release note generation, bug prediction, and software evolution analysis. Also, it can be reproduced using the latest LLM models like GPT, Gemini, etc. In conclusion, the study demonstrates that large language models can be an important tool in automating conventional commit message classification using different prompting strategies.

\bibliographystyle{IEEEtran}
\bibliography{reference}

\end{document}